\documentclass[apj]{emulateapj}
\usepackage{color}

\def\gtorder{\mathrel{\raise.3ex\hbox{$>$}\mkern-14mu
             \lower0.6ex\hbox{$\sim$}}}
\def\ltorder{\mathrel{\raise.3ex\hbox{$<$}\mkern-14mu
             \lower0.6ex\hbox{$\sim$}}}

\usepackage{graphicx}	
\usepackage{amssymb}	
\usepackage{colordvi}
\usepackage{amsmath}
\usepackage{hyperref}

\usepackage{academicons}

\slugcomment{Submitted to AJ}

\shorttitle{TRANSLIENT}

\shortauthors{Springer et al.}

\begin{document}

\title{TRANSLIENT: Detecting Transients Resulting from Point Source Motion or Astrometric Errors}

\author{
Ofer Springer\altaffilmark{1,2},
Eran~O.~Ofek\altaffilmark{1,$\star$},
Barak Zackay\altaffilmark{1},
Ruslan Konno\altaffilmark{1},
Amir Sharon\altaffilmark{1},
Guy Nir\altaffilmark{3},
Adam Rubin\altaffilmark{4},
Asaf Haddad\altaffilmark{1},
Jonathan Friedman\altaffilmark{1},
Leora Schein Lubomirsky\altaffilmark{6},
Iakov Aizenberg\altaffilmark{5},
Alexander Krassilchtchikov\altaffilmark{1},
Avishay Gal-Yam\altaffilmark{1},
}
\altaffiltext{1}{Department of particle physics and astrophysics, Weizmann Institute of Science, 76100 Rehovot, Israel.}
\altaffiltext{2}{Benin School of Computer Science and Engineering, The Hebrew University of Jerusalem, 9190416 Jerusalem, Israel.}
\altaffiltext{3}{Lawrence Berkeley National Laboratory, 1 Cyclotron Road, MS 50B-4206, Berkeley, CA 94720, USA}
\altaffiltext{4}{European Southern Observatory, 85748 Garching bei München, Germany.}
\altaffiltext{5}{QST Financial Group, 4130 Limassol, Cyprus.}
\altaffiltext{6}{Department of chemical and biological physics, Weizmann Institute of Science, 76100 Rehovot, Israel.}
\altaffiltext{$\star$}{Corresponding Author: eran.ofek@weizmann.ac.il}



\begin{abstract}

Detection of moving sources over complicated background is important for several reasons.
First is measuring the astrophysical motion of the source. 
Second is that such motion resulting from
atmospheric scintillation, color refraction, or astrophysical reasons
is a major source of false alarms for image subtraction methods.
We extend the Zackay, Ofek, \& Gal-Yam 
image subtraction formalism to deal with moving sources.
The new method, named \textsc{translient}
(translational transient) detector,
applies hypothesis testing between the hypothesis that the source is stationary and that the source is moving.
It can be used to detect source
motion or to distinguish between stellar variability and motion.
For moving source detection, we show the superiority of \textsc{translient} over the proper image subtraction, using the improvement in the receiver-operating characteristic curve.
We show that in the small translation limit, \textsc{translient} is an optimal detector of point source motion in any direction.
Furthermore, it is numerically stable, fast to calculate, and presented in a closed form.
Efficient transient detection requires both
the proper image subtraction statistics and the \textsc{translient} statistics: when the translient statistic is higher, then the subtraction artifact is likely due to motion.
We test our algorithm both on simulated data 
and on real images obtained by the Large Array Survey Telescope (LAST).
We demonstrate the ability of \textsc{translient} to distinguish between motion and 
variability, which has the potential to reduce the number of false alarms
in transients detection.
We provide the \textsc{translient} implementation in Python and MATLAB.

\end{abstract}


\section{Introduction}
\label{sec:intro}

Detection and measurement of sub-pixel motions of sources in astronomical images 
is required for two major science cases: 
(i) Detecting motion due to any kind of astrometric noise, or astrophysical motion,
that leads to subtraction artifacts seen in image differencing methods,
and in turn, leads to false alarms in transient detection;
(ii) Measuring astrophysical motion,
including proper motion and parallax measurement,
detecting astrometric binaries (e.g., \citealt{Malkov+2012_Masses_VisualBinaries_Catalog}),
astrometric microlensing (e.g., \citealt{Paczynski1998_astrometricMicrolensing};
\citealt{Gould2000_astrometricMicrolensing_SIM};
\citealt{Lu+2016_astrometricMicrolensing_KeckAO};
\citealt{Sahu+2017_astrometricMicrolensing_WD_mass_HST};
\citealt{Ofek2018_NS_AstrometricML}),
lensed quasars (\citealt{Springer+Ofek2021_TimeDelayI_FluxOnly}, \citealt{Springer+Ofek2021_TimeDelayII_FluxAstrometry}),
binary quasars (e.g., \citealt{Liu2015_BinaryQuasars_FromAstrometricCenteroidShifts}, \citealt{Shen+2019_BinaryQuasarsDetection_Astrometry}),
and studying the black hole in the center of our Galaxy
(e.g., \citealt{Ghez+2005ApJ_StarOrbit_BH_GalCenter}; \citealt{Gillessen+2009ApJ_StellarOrbits_BH_GalCen}).

Although point-spread function (PSF) fitting
(e.g., \citealt{Stetson1987_DAOPHOT, Schechter+1993_DoPHOT}) 
is an excellent approach for measuring point source positions, and hence, their motion, 
it is biased in the presence of a complicated background (e.g., in galaxies).
Due to their ability to work in complicated
background regions,  
image subtraction algorithms are the method
of choice for transient detection
(e.g., \citealt{Phillips+1995_ImageSubtraction, Alard+1998_subtraction, 
Bramich2008MNRAS_ImageSubtraction, Zackay+2016_ZOGY_ImageSubtraction}).
Historically, image differencing resulted in subtraction artifacts, non-Gaussian noise, and excess noise 
compared to the expectation.
Some of these problems were solved by the \cite{Zackay+2016_ZOGY_ImageSubtraction} (ZOGY) algorithm,
which provides an optimal, anti-symmetric, fast, with correct noise propagation, and numerically stable 
solution to the transient detection problem.
Indeed, the application of ZOGY results in a reduced amount of false alarms, sometimes
by two orders of magnitude, compared to older methods.

However, some of the assumptions in the derivation of all the existing methods are still not accurate 
(see discussions in \citealt{Zackay+2016_ZOGY_ImageSubtraction}).
One such assumption, which is the subject of this paper, is that the reference image 
and the new image are perfectly registered.
The breakdown of this assumption is yet another leading reason for the presence 
of false alarms in transient detection.
In practice, the breakdown of this assumption is unavoidable for several reasons:
(i) For ground-based observations atmospheric scintillation shifts the position of stars by an amount
typically larger than the measurement error due to the Poisson noise. Furthermore, these shifts 
are correlated within the $<1$\,arcmin-size isoplanatic patch, meaning that in practice 
the stars' positions are shifted, (almost) randomly, with respect to their mean position.
(e.g., \citealt{Lindegren1980_AtmosphericScintilation_GroundBasedAstrometry, Shao+Colavita1992_AtmosphericScintilation_GroundBasedAstrometry, Ofek2019_Astrometry_Code}).
(ii) Color refraction in the atmosphere or in a telescope, along with the variance of stars' colors, 
introduces variations in the relative positions of stars,
as a function of the airmass or the focal plane position, respectively 
(see e.g., \citealt{Zackay+2016_ZOGY_ImageSubtraction}).
(iii) Poisson errors are limiting the accuracy of image registration;
and 
(iv) Proper motion of astrophysical objects. 

\cite{Zackay+2016_ZOGY_ImageSubtraction} suggested two methods to deal with these astrometric errors.
The first method is to propagate the measured (global or regional) astrometric noise of the image
into the image subtraction formula. This approach does not attempt to measure
motion, but only to ignore flux residuals which may be due to shifts in source positions
between the new and the reference images.
\cite{Zackay+2016_ZOGY_ImageSubtraction} demonstrated the
ability of this approach to deal with astrometric errors.
A disadvantage of this method is that it is sub-optimal, and requires knowledge about
the local astrometric errors\footnote{Typically, the astrometric errors obtained from the astrometric solutions are based on the global rms. However, because the isoplanatic scale may be smaller than the typical angular distance between stars, the astrometric noise is mostly uncorrelated and may change as a function of spatial position.}.
If for example, the displacement of a source between the new and reference images is larger than the assumed astrometric noise, the displaced source will likely generate a detection in the subtraction image.
The second suggestion made in \cite{Zackay+2016_ZOGY_ImageSubtraction},
was to model the flux residual due to a specific astrometric shift in the position of a source by 
modifying the PSF of one of the images to account for the astrometric misalignment.
This suggestion has not been derived and implemented as of yet and is the subject of this paper.

In the current work, we derive a formalism for the detection and measurement of a moving point source on a general static background.
Additionally, we prove that in the small translation limit, our detector is an optimal detector of point source motion in any (non-specific) direction.
This property is what essentially enables the employment of the discussed method as a blind detection scheme 
and not as a post-fitting scheme. We also discuss the extent to which small translations can be 
distinguished from local flux changes and find a set of non-degenerate observational parameters for this
task. We refer to the translational transient detection method presented below by the name 
\textsc{translient} and to transient events arising from point source motion as \textit{translients} 
(translational transients). This paper deals with deriving the formalism and testing the new method on
simulated and real data.

The structure of the paper is as follows: In \S\ref{sec:method} we develop a statistical image formation 
model, derive an optimal statistic to perform hypothesis testing in the motion-only scenario, and derive 
the likelihood of observing a particular difference image when both motion and a change of flux are taking
place. The statistical sufficiency of this difference image is discussed in Appendix \ref{sec:D_suff}. In 
\S\ref{sec:sim} we use simulated images to measure the performance of a \textit{translient} detector as
compared to the \textit{proper subtraction} detector at various levels of background noise and translation
size. We additionally use these simulations to assess the ability to differentiate between motion 
and flux variation when translations are smaller than the PSF size.
In \S\ref{sec:RealData} we present some tests of \textit{translient} on real astronomical images, 
in \S\ref{sec:Code} we describe our software implementations, and we conclude in \S\ref{sec:discussion}.

\section{Methods derivation} 
\label{sec:method}

Extending \cite{Zackay+2016_ZOGY_ImageSubtraction}, we derive the formalism needed to detect
and measure point sources moving on a general static background.
In \S\ref{sec:model} we outline the employed statistical model of
formation of astronomical images. 
We start by discussing the case of
a particular motion in \S\ref{sec:motion_particular},
and continue with the case of
a general motion (direction and amplitude)
in \S\ref{sec:motion_any}.
Next, in \S\ref{sec:discrim}
we discuss how one can distinguish between
flux variation and motion,
and in \S\ref{sec:fitting} we present a method to fit the motion 
and flux variation simultaneously.

\subsection{Image formation model}
\label{sec:model} 

Our model for the reference image ($R$) and the new image ($N$) is given by:
\begin{eqnarray}
R = (T + \alpha_r\delta_{\vec{q}})\otimes P_r + \epsilon_r, \label{eq:model_R}\\
N = (T + \alpha_n\delta_{\vec{p}})\otimes P_n + \epsilon_n. \label{eq:model_N}
\end{eqnarray}
Here $\otimes$ denotes convolution, $T$ is a general unknown true background image,
$r$ and $n$ subscripts indicate the reference image and the new image, respectively,
$\alpha_r$ and $\alpha_n$ are the observed point source fluxes, $P_r$ and $P_n$ are the point-spread-functions (PSF), and $\delta_{\vec{q}}$ and $\delta_{\vec{p}}$ denote Dirac delta functions centered at image positions $\vec{q}$ and $\vec{p}$, in the reference and new image, respectively. The additive background noise of the images are $\epsilon_r$ and $\epsilon_n$ and these are assumed to contain independent and identically distributed (i.i.d.) per-pixel Gaussian noise having zero mean and known variances, $\sigma_r^2$ and $\sigma_n^2$, respectively.
This assumption is typically valid since in most cases astronomical images
are in the background-noise-dominated regime,
therefore, $\sigma_{r}$ and $\sigma_{n}$ can be approximated as scalars.
Furthermore, the background is high enough such that the Gaussian noise approximation holds
(see \citealt{Ofek+2018_PoissonNoiseMF} for matched filtering in the Poisson noise case).
We further assume that the images are Nyquist
sampled\footnote{A PSF that has a two-dimensional Gaussian shape, is not band limited. However, in practice, the information content in the edges of the Fourier Transform of the Gaussian is small, and for practical reasons, it is enough to assume that the PSF full-width at half-maximum (FWHM) is about $\gtrsim2.3$ pixels (see e.g., the effect on astrometry in \cite{Ofek2019_Astrometry_Code}).},
and are flux matched\footnote{In the notation of the \cite{Zackay+2016_ZOGY_ImageSubtraction} algorithm this means that $F_{\rm r}=F_{\rm n}=1$.}.

\subsection{Detecting pure motion for particular translation}
\label{sec:motion_particular}

We now specialize our model to the case where we assume that the point source has a constant flux 
($\alpha_r=\alpha_n\equiv\alpha$). At each reference image position $\vec{q}$ we would like our detector to test between the following two hypotheses:
\begin{eqnarray}
\mathcal{H}_0 & : \vec{q}=\vec{p}, \\
\mathcal{H}_1 & : \vec{q}\neq \vec{p},
\end{eqnarray}
where the null hypothesis means that the point source is static in both images $R$ and $N$. Note that a static point source can equivalently be absorbed into the background image $T$. The alternative hypothesis $\mathcal{H}_1$ states that a point source, existing in image $R$ at position $\vec{q}$, has translated to position $\vec{p}$ in the image $N$.\\

Motivated by the Neyman-Pearson
lemma\footnote{The Neyman-Pearson lemma states that the likelihood ratio test is the most powerful method for 
testing a simple hypothesis against an alternative.}
\citep{Neyman+Pearson1933_HypothesisTesing}, 
and following ZOGY, we express the likelihood-ratio of the alternative and the null hypotheses\footnote{As shown 
later, our hypothesis can be regarded as a simple hypothesis.}
\begin{eqnarray}
LR(q;p,\alpha) &=& \frac{P(N,R|\mathcal{H}_1)}{P(N,R|\mathcal{H}_0)} \\ \nonumber
&=& \frac{P(N|R,\mathcal{H}_1)}{P(N|R,\mathcal{H}_0)}\frac{P(R|\mathcal{H}_1)}{P(R|\mathcal{H}_0)} \\ \nonumber
&=& \frac{P(N|R,\mathcal{H}_1)}{P(N|R,\mathcal{H}_0)}, 
\end{eqnarray}
where we have used the law of conditional probability in the second equality; and in the third equality we used the fact that the probability
distribution of $R$ is insensitive to which hypothesis is valid.
We use the following definition of the discrete Fourier transform (DFT) of an $m\times m$ pixel image $f[x,y]$
\begin{equation}
\mathcal{F}[f] = \widehat{f}[k_x, k_y] = \sum_{x=0}^{m-1} \sum_{y=0}^{m-1} f[x,y] \exp\left(-2\pi i\frac{\vec{k}\cdot\vec{q}}{m}\right).
\end{equation}
Here the hat sign marks the Fourier transform operator, the dot sign indicates the dot product,
$m$ is the image size, and $\vec{k}\equiv(k_{\rm x}, k_{\rm y})$ are the frequency coordinates.
Taking the Fourier transform of $R$ and $N$ (applying the convolution theorem)
\begin{eqnarray}
\widehat{R} &=& \left(\widehat{T} + \alpha_r\widehat{\delta}_{\vec{q}}\right) \widehat{P}_r + \widehat{\epsilon}_r, \label{eq:R_hat}\\
\widehat{N} &=& \left(\widehat{T} + \alpha_n\widehat{\delta}_{\vec{p}}\right) \widehat{P}_n + \widehat{\epsilon}_n. \label{eq:N_hat}
\end{eqnarray}
Eliminating $\widehat{T}$ between Equations \ref{eq:R_hat} and \ref{eq:N_hat} we obtain
\begin{equation}
\widehat{N} = \left[\frac{\widehat{R}-\widehat{\epsilon}_r}{\widehat{P}_r} + \left(\alpha_n\widehat{\delta}_{\vec{p}} - \alpha_r\widehat{\delta}_{\vec{q}} \right)\right] \widehat{P}_n + \widehat{\epsilon}_n .
\end{equation}

We now use the fact that the DFT of an $m\times m$ zero mean Gaussian noise image with i.i.d. pixels
each with variance $\sigma^2$, is a complex valued $m\times m$ random variable with i.i.d. real and imaginary frequency coefficients each with variance $m^2\sigma^2/2$. We can therefore express the log-likelihood (up to a constant term
which does~not depend on the data)
of the new frequency-image conditioned on the reference image under $\mathcal{H}_0$ as 
\begin{eqnarray}
\log P(\widehat{N}|\widehat{R},\mathcal{H}_0) = -\sum_{k_x, k_y} \frac{|\widehat{N} - \widehat{R}\widehat{P}_n/\widehat{P}_r|^2}{2\mathrm{Var}(\widehat{\epsilon}_n +  \widehat{\epsilon}_r\widehat{P}_n/\widehat{P}_r)} \\ \nonumber
= -\frac{1}{2m^2} \sum_{k_x, k_y} \frac{|\widehat{P}_r\widehat{N} - \widehat{P}_n \widehat{R}|^2}{|\widehat{P}_r|^2\sigma_n^2 + |\widehat{P}_n|^2\sigma_r^2} \equiv {LL}_0,
\end{eqnarray}
where the second line can be obtained from the first line by multiplying the numerator and denominator by $|\widehat{P}_r|^2$ and using the definition of the variances of the noise terms $\widehat{\epsilon}_r$ and $\widehat{\epsilon}_n$. 

In a similar fashion, for the alternative $\mathcal{H}_1$ and setting $\alpha \equiv \alpha_r = \alpha_n$, we obtain
\begin{eqnarray}
\log P(\widehat{N}|\widehat{R},\mathcal{H}_1) = \qquad\qquad\qquad\qquad\qquad\qquad\quad \\ \nonumber
-\frac{1}{2m^2} \sum_{k_x, k_y} \frac{|\widehat{P}_r\widehat{N} - \widehat{P}_n \widehat{R} + \alpha \widehat{P}_n\widehat{P}_r(\widehat{\delta}_{\vec{q}} - \widehat{\delta}_{\vec{p}})|^2}{|\widehat{P}_r|^2\sigma_n^2 + |\widehat{P}_n|^2\sigma_r^2} \equiv {LL}_1.
\end{eqnarray}
Writing $a \equiv \widehat{P}_r\widehat{N} - \widehat{P}_n \widehat{R}$ and $b \equiv \alpha \widehat{P}_n\widehat{P}_r(\widehat{\delta}_{\vec{q}} - \widehat{\delta}_{\vec{p}})$ we can express the log-likelihood-ratio (up-to a constant) as follows
\begin{equation}
\label{eq:LLR}
LLR = {LL}_1 - {LL}_0 = -\frac{1}{2m^2} \sum_{k_x, k_y} \frac{|b|^2 + 2\Re[a\bar{b}]}{|\widehat{P}_r|^2\sigma_n^2 + |\widehat{P}_n|^2\sigma_r^2},
\end{equation}
where the $|b|^2$ term can be 
dropped because it does~not depend on the observations.
Here, $\Re$ represents the real-part function, and the bar sign above a variable indicates the complex-conjugate
operator. If the bar comes above the hat sign it means
that the complex conjugate operator follows the Fourier transform.
Defining the translation vector $\vec{\Delta} \equiv \vec{p} - \vec{q}$, and the center position $\vec{x}_c$ between point source positions $\vec{p}$ and $\vec{q}$ to be $\vec{x}_c \equiv (x_{c}, y_{c}) \equiv \frac{\vec{p}+\vec{q}}{2}$, we obtain
\begin{eqnarray}
\label{eq:delta_p_delta_q}
\widehat{\delta}_{\vec{q}} - \widehat{\delta}_{\vec{p}} &=& e^{-2\pi i\vec{k}\cdot\vec{q}/m} - e^{-2\pi i\vec{k}\cdot\vec{p}/m} \\ \nonumber
&=& e^{-2\pi i\vec{k}\cdot\vec{x}_c/m}\left(e^{2\pi i\vec{k}\cdot\vec{\Delta}/2m} - e^{-2\pi i\vec{k}\cdot\vec{\Delta}/2m}\right) \\ \nonumber
&\approx& e^{-2\pi i\vec{k}\cdot\vec{x}_c/m}\left[
   \left(1 + \frac{\pi i\vec{k}\cdot\vec{\Delta}}{m} \right) - 
   \left(1 - \frac{\pi i\vec{k}\cdot\vec{\Delta}}{m} \right)\right] \\ \nonumber
&=& \frac{2\pi i \vec{k}\cdot\vec{\Delta}}{m} e^{-2\pi i\vec{k}\cdot\vec{x}_c/m},
\end{eqnarray}
where the approximation between the second and third lines holds for sufficiently small translations where we may keep only terms first order in $\vec{k}\cdot\vec{\Delta}$.

Using the fact that $|b|^2 = \mathcal{O}((\vec{k}\cdot{\vec{\Delta}})^2)$, we find
\begin{eqnarray}
\Re[a\bar{b}] \approx \quad\qquad\qquad\qquad\qquad\qquad\qquad\qquad\qquad\\ \nonumber
\frac{2\pi \alpha \vec{k}\cdot\vec{\Delta}}{m}\Im \left[\bar{\widehat{P}}_n \bar{\widehat{P}}_r(\widehat{P}_r\widehat{N} - \widehat{P}_n \widehat{R}) e^{2\pi i\vec{k}\cdot\vec{x}_c/m}\right], 
\end{eqnarray}
where the $\Im$ function
is the imaginary-part function that replaces the $\Re$ function due to multiplication by the imaginary unit $i$. Substituting the above into the expression for the $LLR$ (Equation \ref{eq:LLR}) we find that
\begin{eqnarray}
LLR \approx -\frac{1}{2m^2} \sum_{k_x, k_y} \frac{2\Re[a\bar{b}]}{|\widehat{P}_r|^2\sigma_n^2 + |\widehat{P}_n|^2\sigma_r^2} = \qquad\qquad\qquad\qquad\\ \nonumber
\alpha\vec{\Delta}\cdot\Im\left\{\frac{1}{2m^2}\sum_{k_x, k_y} \left[\frac{4\pi\vec{k}}{m}\frac{\bar{\widehat{P}}_n \bar{\widehat{P}}_r(\widehat{P}_n\widehat{R} - \widehat{P}_r \widehat{N})}{|\widehat{P}_r|^2\sigma_n^2 + |\widehat{P}_n|^2\sigma_r^2} e^{2\pi i\vec{k}\cdot\vec{x}_c/m}\right]\right\}.   
\end{eqnarray}

Defining the following two component frequency image (one component for each spatial frequency direction) 
\begin{equation}
\vec{\widehat{z}}\equiv \frac{4\pi\vec{k}}{m}\frac{\bar{\widehat{P}}_n \bar{\widehat{P}}_r(\widehat{P}_n\widehat{R} - \widehat{P}_r \widehat{N})}{|\widehat{P}_r|^2\sigma_n^2 + |\widehat{P}_n|^2\sigma_r^2},
\end{equation}
and identifying the inverse DFT, with respect to the image coordinate $\vec{x}_c$, in the last expression for the $LLR$, we define
\begin{equation}
\vec{z}\equiv \mathcal{F}^{-1} \left[ \vec{\widehat{z}}\right],
\end{equation}
and find (to first order in $\vec{k}\cdot\vec{\Delta}$) that
\begin{eqnarray}
LLR(q;\alpha\vec{\Delta}) &\approx& \alpha\vec{\Delta}\cdot\Im\left\{\mathcal{F}^{-1}\left[\vec{\widehat{z}}\right]\right\} \\ \nonumber
&=& \alpha\vec{\Delta}\cdot\mathrm{Im}\left[\vec{z}\right]=\alpha\Delta\left(\vec{\Delta}_u\cdot\mathrm{Im}\left[\vec{z}\right]\right),
\end{eqnarray}
where we use the symbol $\vec{\Delta}_u \equiv \frac{\Delta}{||\vec{\Delta}||}$ to denote the unit vector in the direction of the translation of $\vec{\Delta}$.

Then, according to the Neyman-Pearson lemma, rejecting $\mathcal{H}_0$ when $\vec{\Delta}_u\cdot\Im\left[\vec{z}\right] > \eta$, for some threshold $\eta$, is the most powerful test at significance level $P(\vec{\Delta}_u\cdot\Im\left[\vec{z}\right] > \eta | \mathcal{H}_0)$. This assumes that the direction of translation, $\vec{\Delta}_u$ is known. We generalize this in the following section.
\\

\subsection{Detecting pure motion for any translation}
\label{sec:motion_any}

To find a statistic that is most powerful at a certain significance level for any direction of translation, we modify the alternative hypothesis of the previous section to the following
\begin{equation}
\tilde{\mathcal{H}}_1 : ||\vec{q}-\vec{p}|| = \Delta. \\
\end{equation}
This leads to the following likelihood ratio: 
\begin{equation}
LR(q;\Delta,\alpha) = \frac{P(N,R|\tilde{\mathcal{H}}_1)}{P(N,R|\mathcal{H}_0)}.
\end{equation}
Defining $\theta$ to be the angle between $\vec{\Delta}_u$ and the positive $x$-axis in the image plane, then
\begin{eqnarray}
LR(q;\Delta,\alpha) = \int_{0}^{2\pi} d\theta \exp\left(LLR(q;\alpha\vec{\Delta})\right) \\ \nonumber
= \int_{0}^{2\pi} d\theta' \exp\left(
\alpha\Delta\left(\cos{\theta'} ||\Im\left[\vec{z}\right]||\right)\right),
\end{eqnarray}
which, for any positive value of $\alpha\Delta$, is an increasing function of $||\Im\left[\vec{z}\right]||$ (this can be shown using the fact that for every $r > 0$, $\frac{d}{dr}\left[\exp(r) + \exp(-r)\right] = \exp(r) - \exp(-r) > 0$).

We thus conclude that a threshold test of the statistic 
\begin{eqnarray}
Z^2(x, y) &\equiv& ||\Im[\vec{z}]||^2 \label{eq:translient_stat}
 \\ \nonumber
&=&
\left|\left|
\Im
\left\{ \mathcal{F}^{-1} \left[
    \frac{4\pi\vec{k}}{m}\frac{\bar{\widehat{P}}_n \bar{\widehat{P}}_r(\widehat{P}_n\widehat{R} - \widehat{P}_r \widehat{N})}{|\widehat{P}_r|^2\sigma_n^2 + |\widehat{P}_n|^2\sigma_r^2}
\right] \right\} 
\right|\right|^2,
\end{eqnarray}
is a most powerful test, at a specific significance level, for detecting a point source, having a center pixel 
coordinates $(x,y)\equiv\vec{x}_c\equiv (\vec{p}+\vec{q})/2$
undergoing a translation in \textit{any} direction. 
We call this last expression the {\it translient statistic} ($Z^{2}$).
An important property of Equation~\ref{eq:translient_stat} is that it is anti-symmetric\footnote{Meaning that one can switch the role of $R$ and $N$ and this will return the same result with opposite sign.} and numerically stable\footnote{The numerator converges to zero faster than the denominator.} (i.e., no division by zero).
Given the definition of $Z^{2}$, its values are distributed like the $\chi^{2}$ distribution with two degrees of freedom.

\subsection{Distinguishing motion from flux variation}
\label{sec:discrim} 
Given a transient detection, we would like to know what the likely cause of the transient detection is: a motion of a point source or a change in its flux.

In the case that we are interested only in knowing which model (flux variation or motion) better explains the 
residuals, we can compare the translient statistic $Z^2$ (Equation \ref{eq:translient_stat}) against the proper 
subtraction statistic $S$, where $S$ is defined in Equations (A26) and (A27) of 
\cite{Zackay+2016_ZOGY_ImageSubtraction}. As the two statistics are strictly not produced by nested models, the 
comparison has to occur indirectly.
We use two equivalent methods:
(i) to convert the $Z^{2}$ to Gaussian significance (assuming $\chi^{2}$ distribution with two degrees of freedom) -- we call this significance $Z_{\sigma}$, and to compare it with $\vert S\vert$;
(ii) Since $S$ follows a normal distribution, $S^{2}$ is distributed like a $\chi^{2}$ distribution with one degree of freedom. Therefore, we can compare $S^{2} + 1$ to $Z^{2}$.

We note that before using 
$S^{2}$ and $Z^{2}$
(or $S$ and $Z_{\sigma}$),
it is recommended to normalize them empirically, such that $S$ will have a mean (or median) of zero and a standard deviation (or robust standard deviation) of 1.
Similarly, $S^{2}$ and $Z^{2}$
can be normalized such that their
mean, median, or variance, will be
 $k$, $k (1 - 2/(9k))^{3}$, or $\sqrt{2k}$, respectively,
where $k$ is the number of degrees of freedom.
Finally, if $Z^{2}>S^{2}+1$ (or $Z_{\sigma}>\vert S\vert$), then the motion model is preferred,
while if $Z^{2}<S^{2}+1$, then the variability model is more likely.

An important implementation comment is that,
as seen in Figure~\ref{fig:residuals_pos_sims_1}, the local maxima in $\vert S\vert$ and $Z_{\sigma}$ are not 
at the same location.
Therefore, our recommendation for implementation is as follows:
(i) Look for a local maximum in $\vert S\vert$ that is larger than the detection threshold (e.g., 5$\sigma$);
(ii) For each local maximum in $\vert S\vert$, look for the maximum value of $Z_{\sigma}$ within a radius of $X$ 
pixels from the local maximum position in $\vert S\vert$;
(iii) If $\vert S\vert > Z_{\sigma}$, then declare that the local maximum is a transient candidate.
In our own implementation which is described below, we used a radius of $X=5$\,pix (about 2--2.5 times the FWHM).
When we compare $\vert S\vert$ and $Z_{\sigma}$, we always refer to the maximum of $Z_{\sigma}$ found near the local maximum of $\vert S\vert$.

\subsection{Fitting motion and flux variation}
\label{sec:fitting}

Another approach is to fit a model that mixes flux variations and motion,
and therefore allows one to consider more complicated cases, in which the transient is both moving and variable.
To do so, we use the new and reference image models defined in equations \ref{eq:model_R} and \ref{eq:model_N}, 
where we now allow for observed flux variation, so that $\alpha_r$ will generally differ from $\alpha_n$.
Using the definitions of the Fourier transforms of the reference image $\widehat{R}$ and the new image 
$\widehat{N}$  (equations \ref{eq:R_hat} and \ref{eq:N_hat}), we may eliminate $\widehat{T}$ by writing
\begin{equation}
    \widehat{P}_r \widehat{N} - \widehat{P}_n \widehat{R} = 
    \widehat{P}_n\widehat{P}_r \left( 
        \alpha_n\widehat{\delta}_{\vec{p}} - 
        \alpha_r\widehat{\delta}_{\vec{q}} \right) + 
        \widehat{P}_r\widehat{\epsilon}_n - \widehat{P}_n\widehat{\epsilon}_r.
\end{equation}
This motivates the introduction of the following frequency domain difference image:
\begin{eqnarray}
\widehat{D}_T &\equiv& \widehat{P}_r\widehat{N} - \widehat{P}_n \widehat{R} + \widehat{P}_n\widehat{P}_r(\alpha_r\widehat{\delta}_{\vec{q}} - \alpha_n\widehat{\delta}_{\vec{p}}) \label{eq:D} \\ \nonumber
&=& \widehat{P}_r\widehat{\epsilon}_n - \widehat{P}_n\widehat{\epsilon}_r.
\end{eqnarray}
Here, we get the second line by using equations~\ref{eq:model_N} and \ref{eq:model_R}.
In Appendix \ref{sec:D_suff} we show that this specific choice of a difference image $\widehat{D}_T$ captures all 
the available information (from the observed $R$ and $N$) for the purpose of finding the model parameters 
$\alpha_r$, $\alpha_n$, $\vec{q}$, and $\vec{p}$. In other words, that $\widehat{D}_T$ is a \textit{sufficient 
statistic}\footnote{In statistics, a statistic is sufficient with respect to a statistical model and its associated 
unknown parameter if no other statistic that can be calculated from the same sample provides any additional 
information as to the value of the parameter.}
with respect to the model parameters.

We now express the log-likelihood of observing $\widehat{D}_T$ for a given $\alpha_r$, $\alpha_n$, $\vec{q}$ 
and $\vec{p}=\vec{q}+\vec{\Delta}$ as
\begin{eqnarray}
\label{eq:log_P_D}
\log P(\widehat{D}_T | \alpha_r, \alpha_n, \vec{q}, \vec{p}) = -\log \zeta\quad\qquad\qquad\qquad\qquad\\ \nonumber
-\frac{1}{2m^2} \sum_{k_x, k_y} \frac{|\widehat{P}_r\widehat{N} - \widehat{P}_n \widehat{R} + \widehat{P}_n
\widehat{P}_r(\alpha_r\widehat{\delta}_{\vec{q}} - \alpha_n\widehat{\delta}_{\vec{p}})|^2}{|\widehat{P}_r|^2
\sigma_n^2 + |\widehat{P}_n|^2\sigma_r^2},
\end{eqnarray}
where we have included the Gaussian normalization term
\begin{equation}
\log \zeta \equiv 
\frac{1}{2}\sum_{k_x,k_y}\log\left[2\pi m^2\left( 
    |\widehat{P}_r|^2\sigma_n^2 + |\widehat{P}_n|^2\sigma_r^2 \right)\right].
\end{equation}
We can specialize Equation~\ref{eq:log_P_D} to the small translation limit by keeping only terms first order in 
$\vec{k}\cdot\vec{\Delta}$. This can be achieved by substituting:
\begin{eqnarray}
\alpha_r\widehat{\delta}_{\vec{q}} - \alpha_n\widehat{\delta}_{\vec{p}} &=& 
    \alpha_r e^{-2\pi i\frac{\vec{k}\cdot\vec{q}}{m}} -  
    \alpha_n e^{-2\pi i\frac{\vec{k}\cdot\vec{p}}{m}} \\ \nonumber
&=& e^{-2\pi i\vec{k}\cdot\frac{\vec{q}+\vec{p}}{2m}} \left(
\alpha_r e^{2\pi i\vec{k}\cdot\frac{\vec{\Delta}}{2m}} - 
\alpha_n e^{-2\pi i\vec{k}\cdot\frac{\vec{\Delta}}{2m}} \right) \\ \nonumber
&\approx& e^{-2\pi i\frac{\vec{k}\cdot\vec{x}_c}{m}} \left[(\alpha_r-\alpha_n) + (\alpha_r+\alpha_n)\frac{\pi i (\vec{k}\cdot\vec{\Delta})}{m}\right],
\label{eq:alpha_r_delta_q_minus_alpha_n_delta_p}
\end{eqnarray}
where we again use the image coordinate $\vec{x}_c\equiv\frac{\vec{q}+\vec{p}}{2}$. In this limit, a set of 
independent parameters for the likelihood are the flux difference $(\alpha_r - \alpha_n)$, the flux sum multiplied 
by the magnitude of translation $(\alpha_r + \alpha_n)||\Delta||$ and the angle $\theta$ between the direction of 
translation $\vec{\Delta}_u$ and the positive direction of the $x$-axis. 

We also write Equation~\ref{eq:log_P_D} using the following two alternative parametrizations 
(which we will find useful in \S\ref{sec:sim_discrim}):
\begin{eqnarray}
\log P(\widehat{D}_T | \alpha_r, \alpha_n, \vec{\Delta}, \vec{x}_c) = \label{eq:P_DT_Delta} \\
\log P(\widehat{D}_T | \alpha_r-\alpha_n, \alpha_r+\alpha_n, ||\Delta||, \theta, \vec{x}_c) = \label{eq:P_DT_4_param} \\ 
\log P(\widehat{D}_T | \alpha_r, \alpha_n, \vec{q}, \vec{p}), \nonumber
\end{eqnarray}
and we denote the linearized version of Equation~\ref{eq:log_P_D} where we substitute the above approximation for 
$\alpha_r\widehat{\delta}_{\vec{q}} - \alpha_n\widehat{\delta}_{\vec{p}}$ (the left hand side of Equation 
\ref{eq:alpha_r_delta_q_minus_alpha_n_delta_p}) as
\begin{eqnarray}
\label{eq:PDT}
\log P(\widehat{D}_{T} | \alpha_r-\alpha_n, (\alpha_r+\alpha_n)||\Delta||, \theta) = \label{eq:P_DT_3_param} \\
\log P(\widehat{D}_T | \alpha_r, \alpha_n, \vec{q}, \vec{p}) +\mathcal{O}(||\vec{\Delta}||^2) \nonumber
\end{eqnarray}
The importance of $D_{\rm T}$ is that it can be used to fit $\alpha_{r}$, $\alpha_{n}$,
$\vec{\Delta}$, $\vec{x}_{c}$, simultaneously (see \S\ref{sec:sim_discrim}).

A disadvantage of this approach over the test suggested in \S\ref{sec:discrim} is that if the source flux is 
constant, fitting the full model will result in lower sensitivity, compared to employment of $Z^2$.
Furthermore, fitting the likelihood in Equation~\ref{eq:PDT} is computationally expensive.
Therefore, we suggest using this approach only in the rare cases 
when both flux variations and motion are suspected.

\section{Simulations}
\label{sec:sim}

In this section, we demonstrate the operation and measure the performance of the translient statistic $Z^2(x,y)$ 
(Equation~\ref{eq:translient_stat}), as well as of the difference image likelihood function $\log P(\widehat{D}_T | 
\alpha_r, \alpha_n, \vec{q}, \vec{p})$ (Equation~\ref{eq:log_P_D}). Specifically, we measure the performance of 
$Z^2(x,y)$ as a detector of pure motion in \S\ref{sec:perf} and the ability of $\log P(\widehat{D}_T | \alpha_r, 
\alpha_n, \vec{q}, \vec{p})$ to distinguish between motion and flux variation in \S\ref{sec:sim_discrim}. 

The performance of the new algorithm is shown on simulated $64\times 64$ pixel image pairs, $R(x,y)$ and $N(x,y)$, 
the synthetic reference, and new images, respectively. These images simulate the generative process of equations 
\ref{eq:model_R} and \ref{eq:model_N} assuming a static true background image (setting $T=0$).
Image pairs are generated using 2-D Gaussian profile point spread functions $P_r(x,y)$ and $P_n(x,y)$ having 
various aspect ratios and orientations (defined below).
To generate $R$ and $N$ with sub-pixel point source positions $\vec{q}$ and $\vec{p}$ we first evaluate the 
(continuous) 2-D Gaussian profiles used to generate $P_r$ and $P_n$ at sub-pixel offsets and then add zero mean 
per pixel Gaussian noise with variances $\sigma_r^2$ and $\sigma_n^2$ respectively. 
In Figure~\ref{fig:stamps_sims_1} we show an example of one such image pair.

\begin{deluxetable}{cccccccc}
\tablecolumns{8}
\tablewidth{0pt}
\tablecaption{Parameters common to detector performance simulations}
\tablehead{
\colhead{$\alpha_r$} & 
\colhead{$\alpha_n$} & 
\colhead{$W_{r}$} & 
\colhead{$W_{n}$} & 
\colhead{$H_{r}$} & 
\colhead{$H_{n}$} & 
\colhead{$\beta_{r}$} & 
\colhead{$\beta_{n}$} \\ 
\colhead{} & 
\colhead{} & 
\colhead{(pix)} & 
\colhead{(pix)} & 
\colhead{(pix)} & 
\colhead{(pix)} & 
\colhead{(deg)} & 
\colhead{(deg)}
} 
\startdata
2.5 & 2.5 & 5.0 & 5.0 & 3.0 & 3.0 & 90 & 0 \\
\enddata
\label{tbl:sim_common_params}
\tablecomments{Here $\alpha_r$ ($\alpha_n$) is the flux level of the point source in the reference (new) image, $W_r$ and $H_r$ ($W_n$ and $H_n$) are the Gaussian profile width and height parameters of the reference (new) image PSF in pixels and $\beta_r$ ($\beta_n$) is the orientation angle of the PSF in the reference (new) image.} 
\end{deluxetable}

\begin{deluxetable}{ccccc}
\tablecolumns{5}
\tablewidth{0pt}
\tablecaption{Noise and translation levels in detector performance simulations}
\tablehead{
\colhead{Simulation} & 
\colhead{$\sigma_r$} & 
\colhead{$\sigma_n$} & 
\colhead{$\Delta_x$} & 
\colhead{$\Delta_y$} \\ 
\colhead{} & 
\colhead{} & 
\colhead{} & 
\colhead{(pix)} & 
\colhead{(pix)}
} 
\startdata
1 & 0.002 & 0.002 & 1.0 & 1.0 \\
2 & 0.003 & 0.003 & 1.0 & 1.0 \\
3 & 0.004 & 0.004 & 1.0 & 1.0 \\
4 & 0.003 & 0.003 & 1.4 & 1.4 \\
5 & 0.003 & 0.003 & 1.1 & 1.1 \\
6 & 0.003 & 0.003 & 0.8 & 0.8 \\
\enddata
\label{tbl:sim_noise_trans_params}
\tablecomments{Here for each simulation, $\sigma_r$ ($\sigma_n$) is the standard deviation value of the per-pixel 
additive white Gaussian flux noise in the reference (new) image $R$ ($N$), measured in the $\alpha_r$ ($\alpha_n$) 
arbitrary flux units. The components of the translation vector $(\Delta_x, \Delta_y) \equiv \vec{\Delta} \equiv 
\vec{p} - \vec{q}$ between the position of the point source in the reference and new images, are in the units of 
pixels. Simulations 1-3 vary the flux noise at constant translation and simulations 4-6 vary the length of the 
translation vector at constant noise levels.}
\end{deluxetable}

\subsection{Performance evaluation of the translient detector}
\label{sec:perf}

\begin{figure}
\vspace{0.33cm}
\centerline{\includegraphics[trim={1.5cm 0 0 0},clip,width=0.48\textwidth]{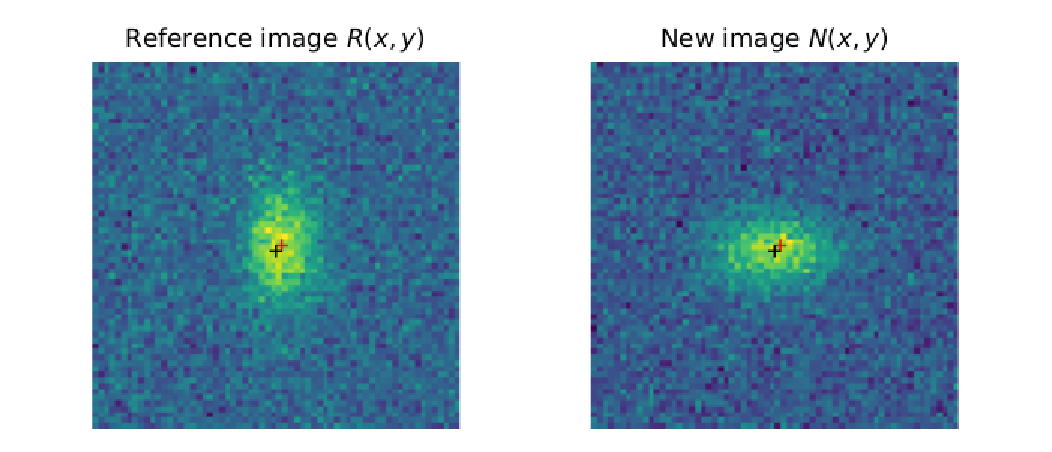}}
\caption{Example of a pair of reference and new images generated in simulation 1 (see 
Table~\ref{tbl:sim_noise_trans_params}). The red and black crosses mark the positions of the original point 
sources in the reference and new images respectively. It can be seen that the translation is a one-pixel 
translation (diagonally), which is considerably smaller than the width of the PSFs and that the Gaussian PSF 
profile of the new image differs from that of the reference image by a $90^{\circ}$ rotation.}
\label{fig:stamps_sims_1}
\end{figure}

We now test the performance of the \textit{translient} statistic and compare it to the \textit{proper image 
subtraction} statistic. Proper image subtraction was designed to be an optimal detector of a general local change 
in flux and does not assume that the transient is due to a translating point source. We thus expect that the 
optimality of the translient statistic in this specific translational data generation process would lead to 
enhanced detector performance. For this purpose, we generate a set of six simulations having varying background 
noise levels (simulations 1-3) and translation sizes (simulations 4-6). The full set of parameters used in these 
simulations is summarized in Tables~\ref{tbl:sim_common_params} and \ref{tbl:sim_noise_trans_params}. In each 
simulation, two sets of $10^4$ image pairs were generated -- one set was of positive examples that did contain a 
translating point source (such as the pair shown in Figure~\ref{fig:stamps_sims_1}) and the other set was of 
negative examples which contained only background noise.

\begin{figure}
\vspace{0.33cm}
\centerline{\includegraphics[trim={1.5cm 0 0 0},clip,width=0.48\textwidth]{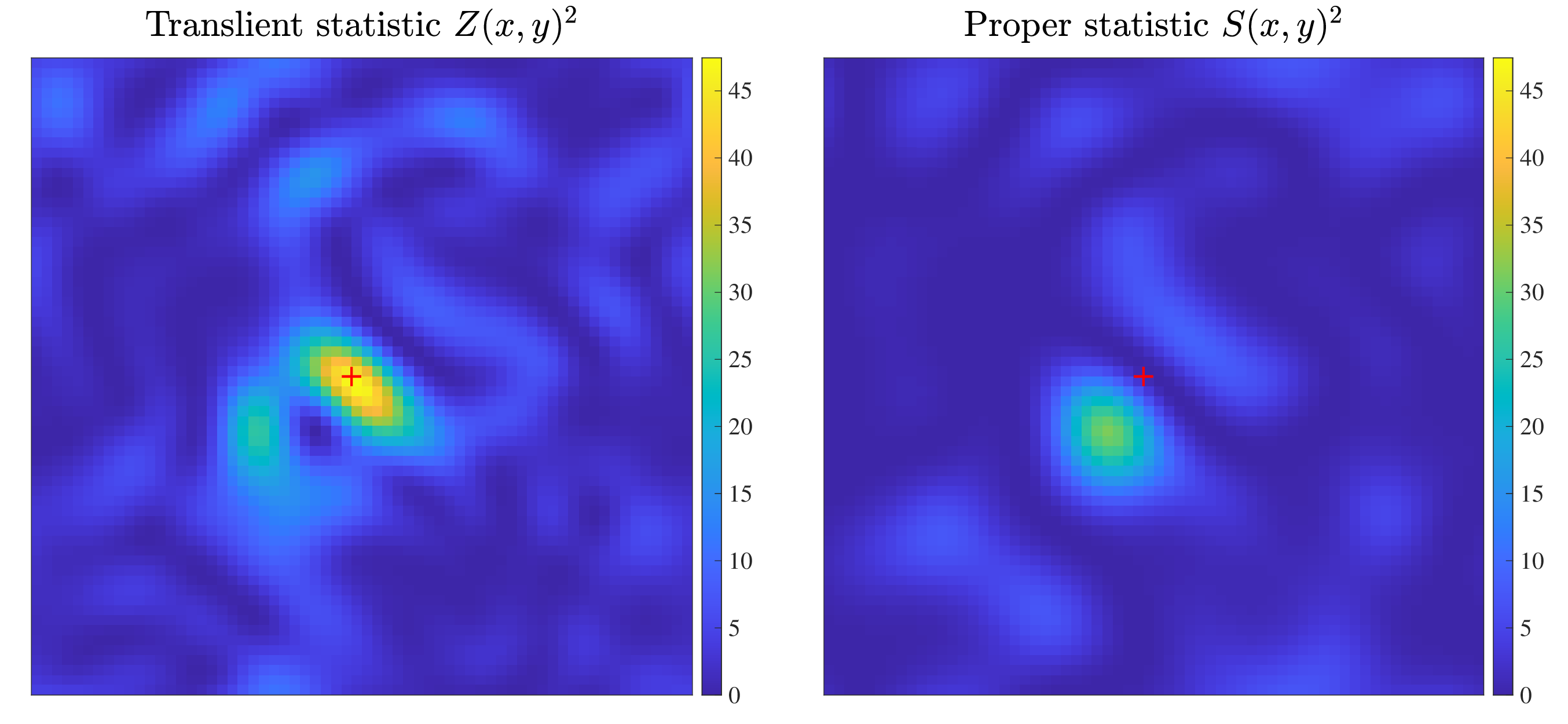}}
\caption{Resulting \textit{translient statistic} ($Z^{2}$; left panel) and \textit{proper statistic} ($S^{2}$; 
right panel) images computed for the image pair shown in Figure~\ref{fig:stamps_sims_1}. 
The red crosses mark the location of $\vec{x}_c \equiv (\vec{p} + \vec{q}) / 2$, the center between the positions of the point sources in the reference and new images. 
In the squared \textit{translient statistic} image one can see a single peak approximately centered at $\vec{x}_c$ 
whereas in the squared \textit{proper statistic} image two peaks on opposite sides of $\vec{x}_c$ (diagonally) can 
be observed. This is the typical qualitative appearance of the two detector images when the data is generated by a 
point source undergoing pure translation (without any flux change).}
\label{fig:residuals_pos_sims_1}
\end{figure}

\begin{figure}
\vspace{0.33cm}
\centering
\includegraphics[trim={1.5cm 0 0 0},clip,width=0.48\textwidth]{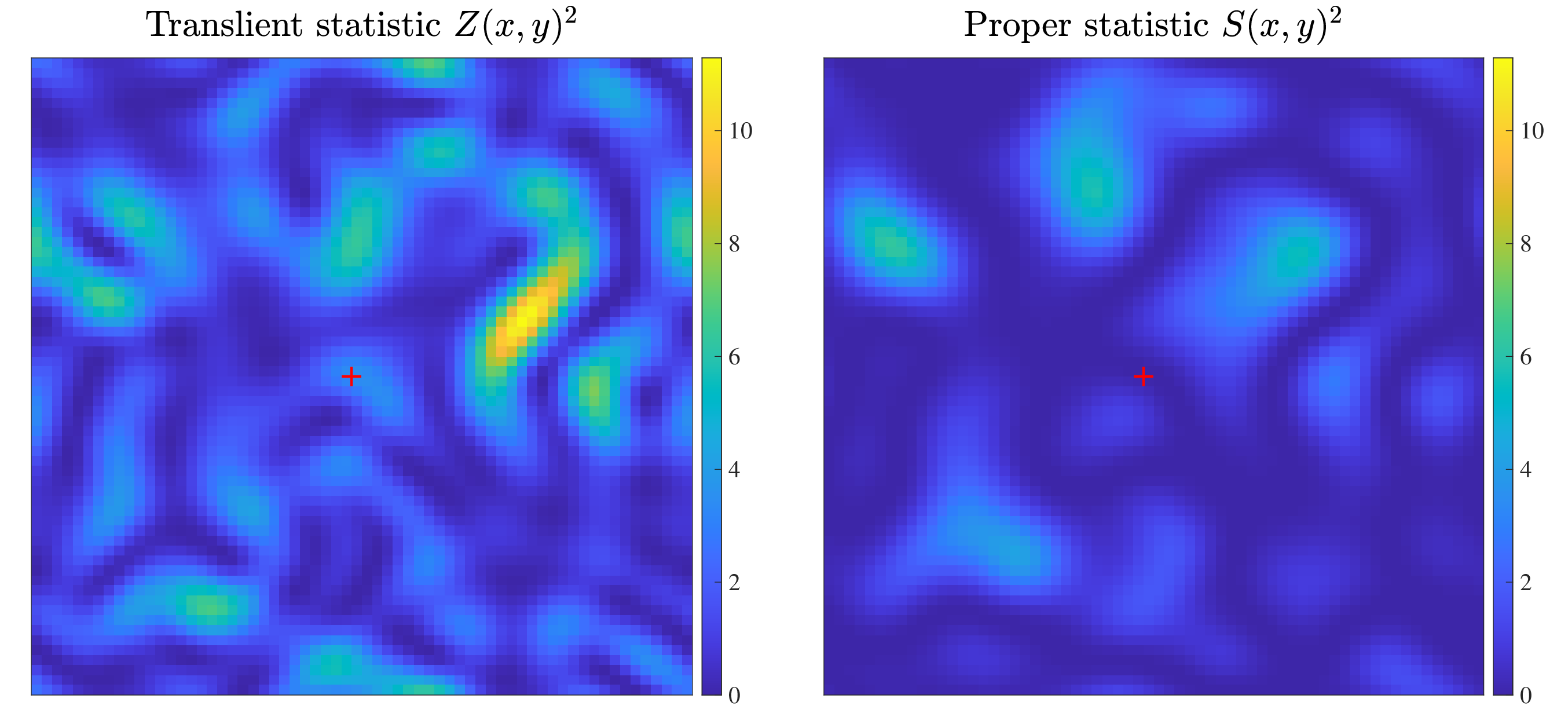}
\caption{Resulting \textit{translient statistic} ($Z^{2}$; left panel) and \textit{proper statistic} ($S^{2}$; 
right panel) images computed for zero-mean noise-only images (no moving point source). These patterns (arising 
from the action of the \textit{translient statistic} and \textit{proper statistic} on the observational noise) are 
responsible for false-positive detection events (at a certain detection threshold), i.e., cases of an event 
being classified as a detection of a translating point source when there was none. 
The highest peak, on the left panel, ($Z^{2}\cong11$) corresponds to a false alarm probability of $0.004$ per 
pixel (i.e., $\chi^{2}$ distribution with two degrees of freedom), and close to unity for $64^{2}$ trials.}
\label{fig:residuals_neg_sims_1}
\end{figure}

For each image pair, we computed the translient statistic image $Z^2(x,y)$ as well as the (squared) proper 
statistic image $S^2(x,y)$.
We show in 
Figure~\ref{fig:residuals_pos_sims_1} an example of these translient and proper statistic images evaluated on the 
positive example of Figure \ref{fig:stamps_sims_1}. We typically see that the center position of the sources 
coincides with the position of the peak in $Z^2(x,y)$ while in $S^2(x,y)$ a double peak (on both sides of the 
source center position) is visible. In Figure~\ref{fig:residuals_neg_sims_1} we show an example of the two image 
statistics when they are evaluated on a negative example pair. These patterns, resulting from the background noise, 
are eventually responsible for the false positive detection events of each of these detectors (note that the 
contrast in Figure~\ref{fig:residuals_neg_sims_1} was enhanced by a factor of ${\sim}5$ compared to that 
of Figure~\ref{fig:residuals_pos_sims_1}).

To quantitatively compare the performance of each image statistic as a detector of pure translation we compute for 
each simulation and for each image statistic (translient and proper) a curve showing the relative rates of the 
true positive detection as a function of the false positive detection rates (known as a receiver operating 
characteristic [ROC] curve). We define the true positive detection rate of each detector (above some 
threshold) to be the number of positive examples that were detected as such, divided by the total number of 
positive examples in the simulation. Similarly, the false positive rate is the number of negative 
examples that the detector incorrectly classified as positive examples divided by the total number of negative 
examples.

\begin{figure}
\vspace{0.33cm}
\centering
\includegraphics[trim={0.25cm 0 0 0},clip,width=0.45\textwidth]{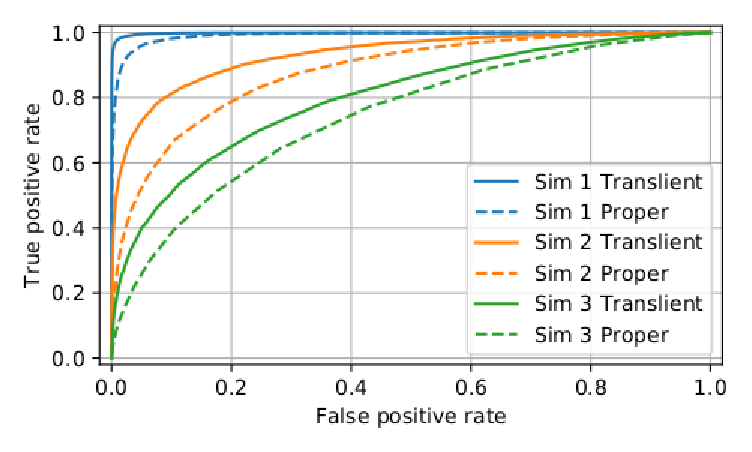}
\caption{\textit{Receiver operating characteristic curve (ROC curve)} for simulations 1-3 (varying levels of flux
noise) showing the rates of true-positive events as a function of the false-positive event rate. These curves were 
generated by varying detector threshold levels $\eta$ over an appropriately wide range. Each color represents a 
simulation index (noise level) while solid (dashed) curves represent the resulting performance when using the 
\textit{translient statistic} (\textit{proper statistic}). One can see that for all simulations and at all values of 
the false-positive rate, the \textit{translient statistic} achieves a higher true-positive detection rate 
compared to the \textit{proper statistic}. We also see that lower levels of noise allow the detectors to reach 
higher true-positive rates at set false-positive rates.}
\label{fig:roc_sims_1_2_3}
\end{figure}

\begin{figure}
\vspace{0.33cm}
\centering
\includegraphics[trim={0.25cm 0 0 0},clip,width=0.45\textwidth]{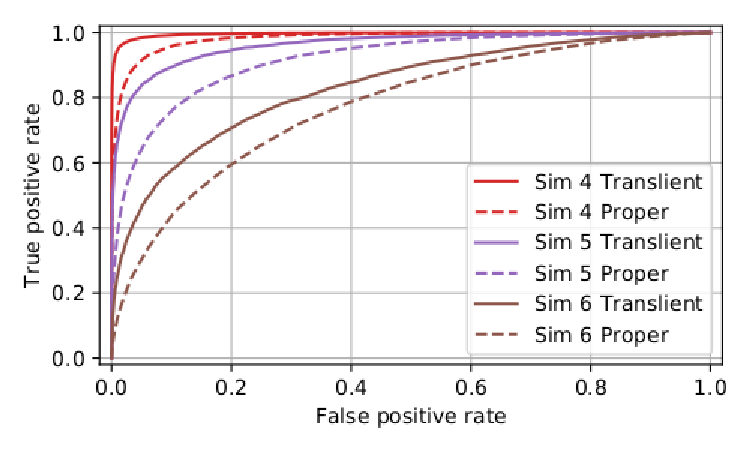}
\caption{ROC curves similar to Figure~\ref{fig:roc_sims_1_2_3} but for simulations 4-6.}
\label{fig:roc_sims_4_5_6}
\end{figure}

For a given threshold level, we define each detector's response to be positive if at least one pixel in the 
statistic image was above this threshold. We show the resulting curves for simulations 1-3 in Figure~\ref{fig:roc_sims_1_2_3} 
(where the noise level was varied),
and for simulations 4-6 in Figure~\ref{fig:roc_sims_4_5_6}
(where the translation length was 
varied).
One can see, that for all simulations and at all values 
of the false-positive rate, that the translient statistic detector achieves a higher true-positive detection 
rate compared to that of the proper statistic detector. One can also see that lower levels of noise and larger 
translations allow both detectors to reach higher true-positive rates for a given set of false-positive rates.

\subsection{Performance evaluation of fitting for motion and flux variation simultaneously}
\label{sec:sim_discrim}

In some rare cases, one may want to check the possibility that the source is both moving and variable.
For this purpose in \S\ref{sec:fitting} we derived expressions for the likelihood of observing the difference 
image $D_T$ due to a point source centered at $\vec{x}_c$ that has translated $\vec{\Delta}$ pixels and changed 
flux from $\alpha_r$ in the reference image to $\alpha_n$ in the new image (Equations \ref{eq:P_DT_Delta}, 
\ref{eq:P_DT_4_param} and \ref{eq:P_DT_3_param}). 
In this section, we use simulations 7 and 8, where both the flux and the position of the point source change, and 
evaluate the likelihood of these observations. The full set of parameters of simulations 7 and 8 is provided in 
Table~\ref{tbl:sim_discrim_params}. Simulation 7 differs from simulation 8 such that in simulation 7 the translation 
is larger than the size of the PSF, whereas in simulation 8 the translation is marginally smaller. 
To evaluate the likelihood functions in this section we perform Markov chain Monte Carlo (MCMC) sampling using the 
\textsc{emcee} package \citep{Foreman-Mackey+2013PASP_EMCMC_Package} by randomly initializing $100$ chains near 
the parameter origin and running each chain through $100$ warm-up iterations and an additional $650$ sampling 
iterations. 

\begin{figure}
\centerline{\includegraphics[trim={0 0 0 0},clip,width=0.45\textwidth]{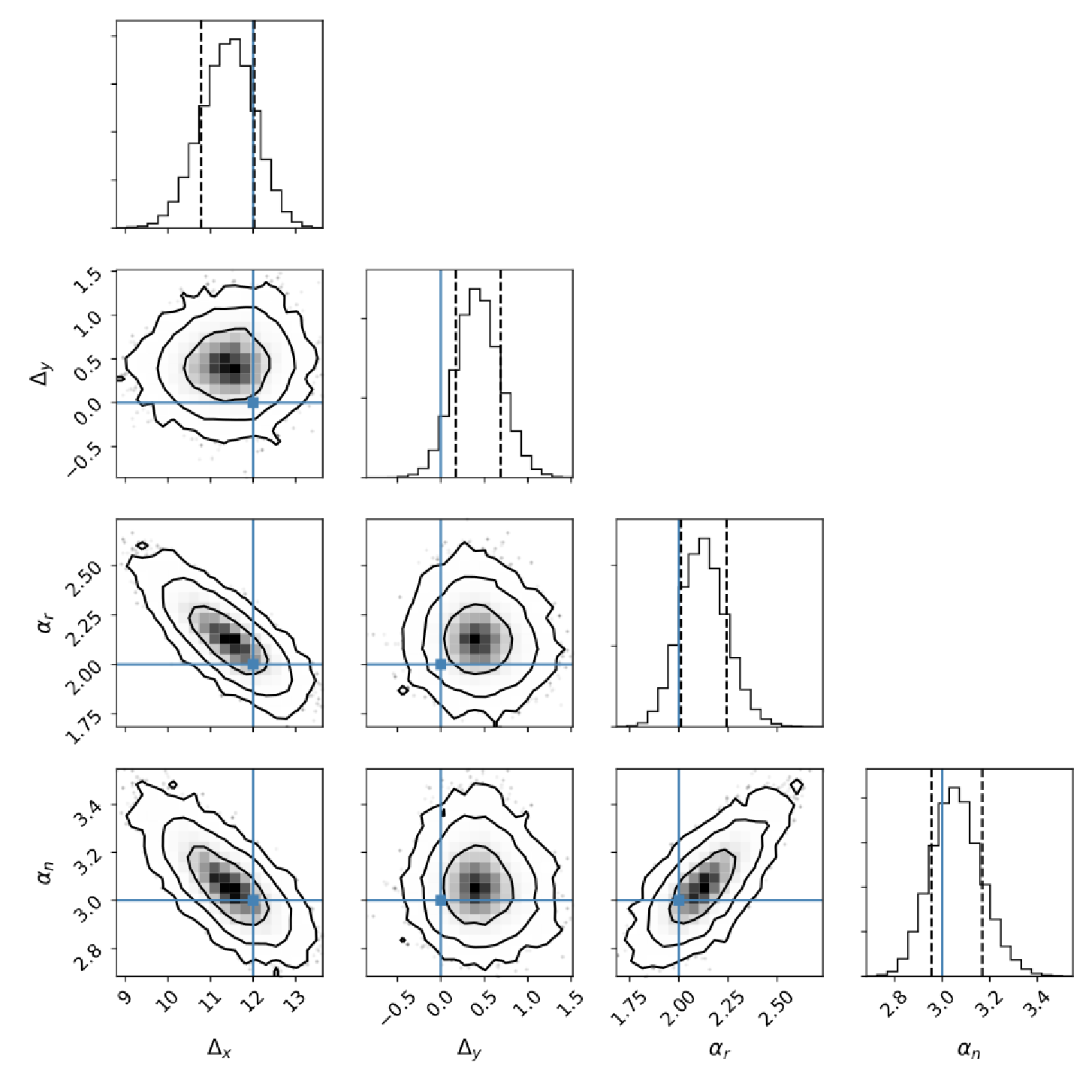}}
  \caption{The likelihood function $P(\widehat{D}_T|\Delta_x, \Delta_y, \alpha_r, \alpha_n)$ evaluated for 
  simulation 7 using the (true) point source center location $\vec{x}_c = (\vec{p}+\vec{q})/2$ as a function of 
  model parameters $\Delta_x$, $\Delta_y$, $\alpha_r$, $\alpha_n$ (see \S\ref{sec:discrim}). The evaluation was 
  performed using Markov chain Monte Carlo sampling. The figure shows the marginal distributions over single 
  parameters (upper diagonal) as well as the marginal distributions for all parameter pairs (off-diagonal). The 
  blue point indicates the true model parameters while the curves indicate the 68\%, 95\% and 99.7\% confidence 
  regions. The dashed vertical lines indicate the 1-sigma confidence intervals in the 1-D marginals.}
\label{fig:discrim_full_non_degenerate}
\end{figure}

We show in Figure~\ref{fig:discrim_full_non_degenerate} 
the resulting MCMC sample distribution of the 
likelihood function $P(\widehat{D}_T | \alpha_r, \alpha_n, \Delta_x, \Delta_y)$
(Equation \ref{eq:P_DT_Delta}, after setting the point source center position $\vec{x}_c$ to its true simulated value), 
when evaluated on simulation 7. In the case of simulation 7, where the point source is well separated between the 
reference and new images (compared to the PSF widths), one can see that the model parameters $\Delta_x$, 
$\Delta_y$, $\alpha_r$, $\alpha_n$ are non-degenerate, and that the resulting sample distribution may be used to 
distinguish between a change in flux and a translation of the point source. 

\begin{figure}
\centerline{\includegraphics[width=0.45\textwidth]{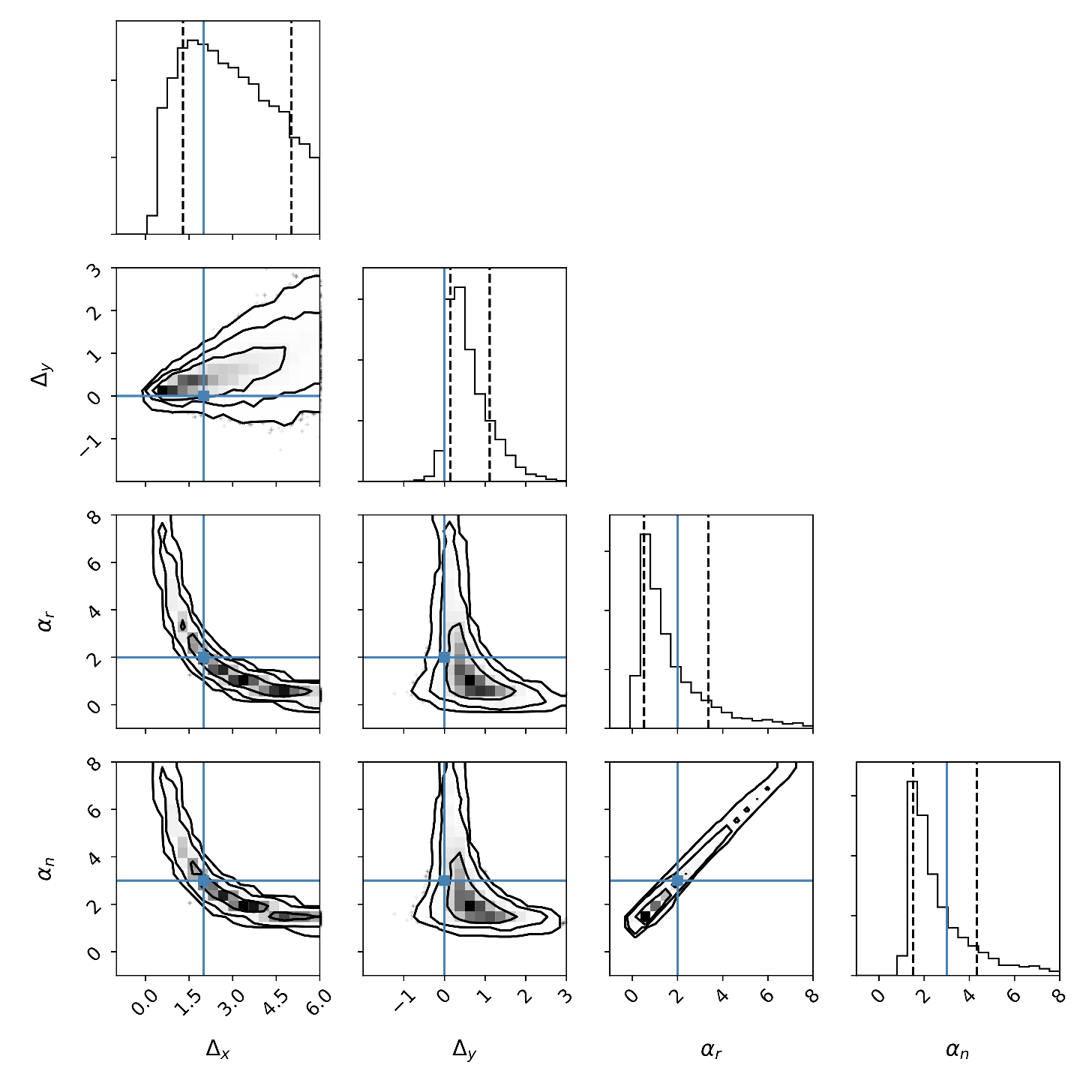}}
  \caption{The likelihood function $P(\widehat{D}_T|\Delta_x, \Delta_y, \alpha_r, \alpha_n)$, similar to 
  Figure~\ref{fig:discrim_full_non_degenerate} but evaluated for simulation 8 where the translation length 
  is smaller than the resolving width of the PSFs.}
\label{fig:discrim_full_degenerate}
\end{figure}

In Figure~\ref{fig:discrim_full_degenerate} we show the resulting MCMC sample distribution for simulation 8, where
the convolved point sources of the reference and new images are not well separated.
One can see that in this case when using model parameters $\Delta_x$, $\Delta_y$, $\alpha_r$, $\alpha_n$, 
there is a strong dependence (or approximate degeneracy) between these parameters. Such a dependence hampers the
ability to distinguish between a change in flux and a translation of the point source.

\begin{figure}
\centerline{\includegraphics[trim={0 0 0 0},clip,width=0.45\textwidth]{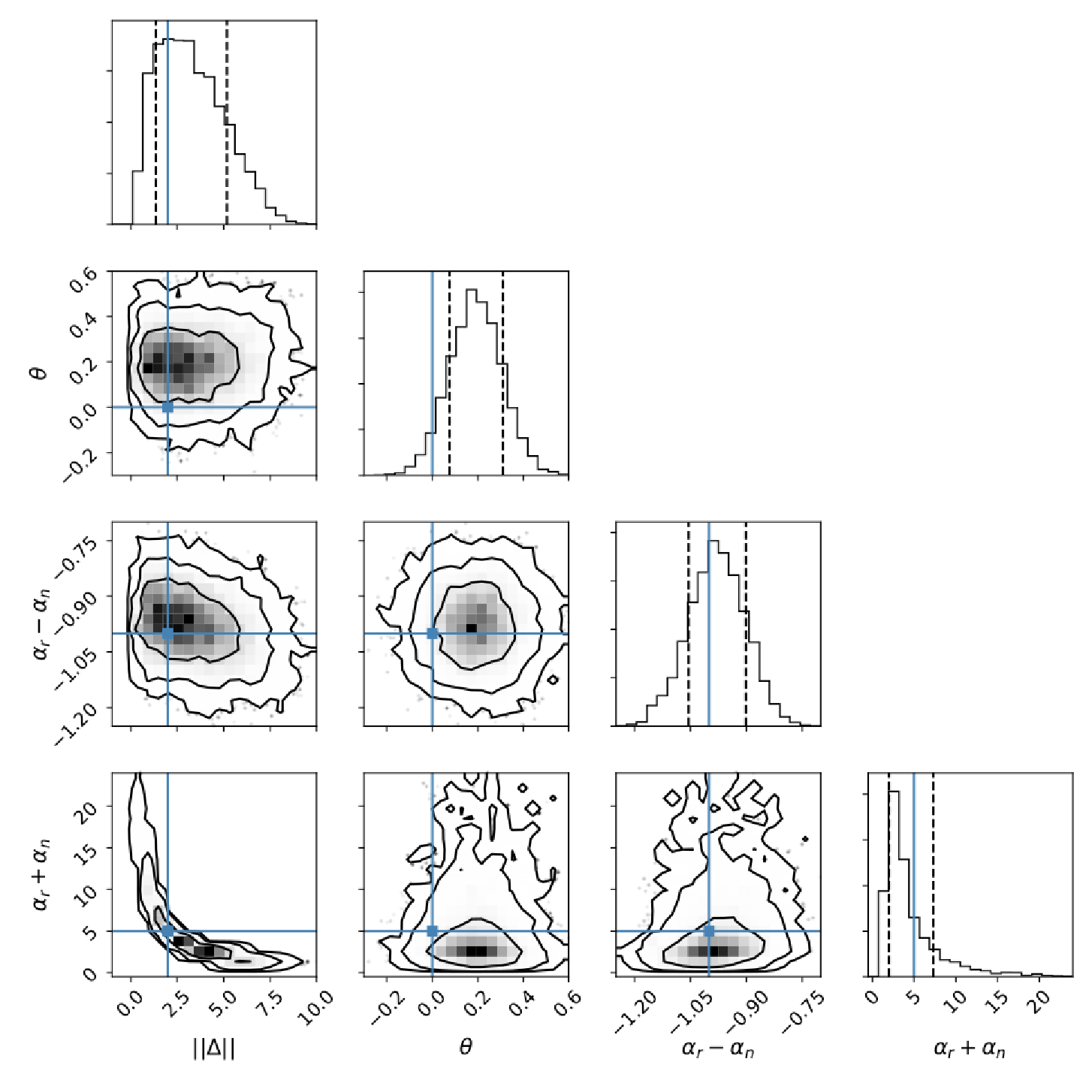}}
  \caption{The likelihood function $P(\widehat{D}_T\,\,|\,\, ||\Delta||, \theta, \alpha_r-\alpha_n, \alpha_r+
  \alpha_n)$, similar to Figure~\ref{fig:discrim_full_degenerate} and similarly evaluated for simulation 8 
  (where the translation is small compared to the PSF) but using a different set of variables to display the same 
  distribution. One can see that this change of variables largely decouples most (though not all) of the parameters, 
  leaving a clear dependence between $||\Delta||$ and $\alpha_r+\alpha_n$.}
\label{fig:discrim_full_degenerate_reparam}
\end{figure}

To partially deal with this degeneracy, one can switch to a new set of (more orthogonal) parameters: $||\Delta||$, $\theta$, $\alpha_r-\alpha_n$ and $\alpha_r+
\alpha_n$.
In Figure~\ref{fig:discrim_full_degenerate_reparam} we present the result
of sampling the same distribution (for simulation 8),
but with this new set of parameters.
One can see that this change of variables largely decouples most (but not all) of the parameters, leaving a 
clear dependence between $||\Delta||$ and $\alpha_r+\alpha_n$. 

\begin{figure}
  \centerline{\includegraphics[width=0.45\textwidth]{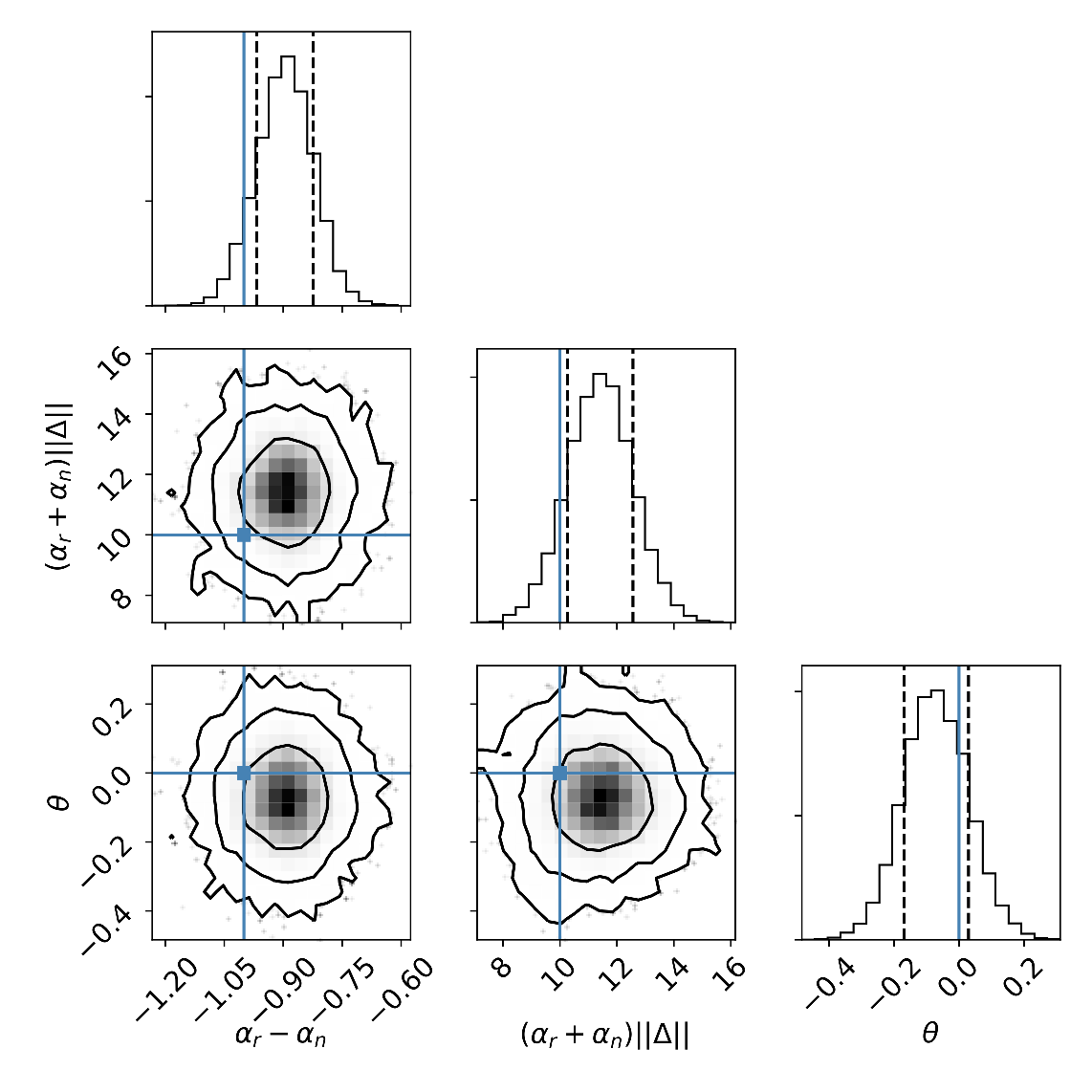}}
  \caption{The likelihood function $P(\widehat{D}_T|\alpha_r-\alpha_n, (\alpha_r+\alpha_n)||\Delta||, \theta)$, 
  obtained by linearizing Equation \ref{eq:log_P_D} in $\vec{k}\cdot\vec{\Delta}$. Here, evaluation is again for 
  the small translation simulation 8 (as in Figures~\ref{fig:discrim_full_degenerate} and 
  \ref{fig:discrim_full_degenerate_reparam}). One can see that this reduced set of model parameters (in the 
  linearized likelihood) does not show a strong dependence and may therefore be preferable in the small translation 
  limit.}
\label{fig:discrim_reduced_degenerate} 
\end{figure}

Finally, we show in Figure \ref{fig:discrim_reduced_degenerate} the result of sampling the linearized likelihood 
(Equation \ref{eq:P_DT_3_param}), again for the ``small translation'' simulation 8. The linearized likelihood depends 
only on the three parameters $\alpha_r-\alpha_n$, $(\alpha_r+\alpha_n)||\Delta||$ and $\theta$ (after setting 
$\vec{x}_c$ to its true simulated value). One can see that when using this reduced set of model parameters, the 
likelihood does not show a strong dependence between the parameters.
To summarize, in some circumstances, inference using this 
set of model parameters can allow us to distinguish motion and flux variation when the translations are smaller than 
the PSF.

\begin{deluxetable}{cccccccc}
\tablecolumns{8}
\tablewidth{0pt}
\tablecaption{Parameters used in the motion vs. flux-variation simulations}
\tablehead{
\colhead{Simulation} & 
\colhead{$\alpha_r$} & 
\colhead{$\alpha_n$} & 
\colhead{$\sigma_r$} & 
\colhead{$\sigma_n$} & 
\colhead{H,W} & 
\colhead{$\Delta_x$} & 
\colhead{$\Delta_y$} \\ 
\colhead{} & 
\colhead{} & 
\colhead{} & 
\colhead{} & 
\colhead{} & 
\colhead{(pix)} & 
\colhead{(pix)} & 
\colhead{(pix)}
} 
\startdata
7 & 2 & 3 & 0.0025 & 0.0025 & 4.0 & 12 & 0 \\
8 & 2 & 3 & 0.0025 & 0.0025 & 4.0 & 2 & 0 \\
\enddata
\label{tbl:sim_discrim_params}
\tablecomments{Parameters of the additional simulations discussed in \S\ref{sec:sim_discrim} to study 
the ability to distinguish between point source motion and point source flux variation. 
The columns are similar to those of Tables~\ref{tbl:sim_common_params} and \ref{tbl:sim_noise_trans_params} 
while here $H$ and $W$ denote the equal horizontal and vertical Gaussian profile diameter of the reference 
and the PSFs of the new image. In both simulations all the parameters other than the translation length are equal. 
In these two simulations, both the flux and the point source position are changed between the reference and new 
images. In simulation 7 the motion is larger than the PSF resolving diameter, while in simulation 8 it is marginally 
smaller than the PSF.}
\end{deluxetable}

\section{Tests on real astronomical images}
\label{sec:RealData}

To test the performance of the translient statistic $Z^2$ (Equation \ref{eq:translient_stat}) on real astronomical
data we have used images obtained with the Large Array Survey Telescope (LAST; \citealt{Ofek+2023PASP_LAST_Overview, 
BenAmi+2023PASP_LAST_Science}). These images have been taken over the course of one night on 2023 Apr 25 
from the LAST site at Neot Smadar, Israel. The data was reduced using the LAST pipeline (\citealt{Ofek2014_MAAT, 
Soumagnac+Ofek2018_catsHTM, Ofek2019_Astrometry_Code, Ofek+2023PASP_LAST_PipeplineI}).  

One of the most important applications of $Z^2$ is to distinguish between flux-variation and motion of a source. 
To test this capability, we derived $Z_{\sigma}$ and $|S|$ on consecutive images of a stationary variable source, 
and of a moving non-variable source.
As explained in \S\ref{sec:discrim}, we first looked for local maxima in $\vert S\vert$,
and then chose the maximum of $Z_{\sigma}$ within 5\,pixels from the local maximum of $\vert S\vert$.

\begin{figure}
\centerline{\includegraphics[width=0.5\textwidth]{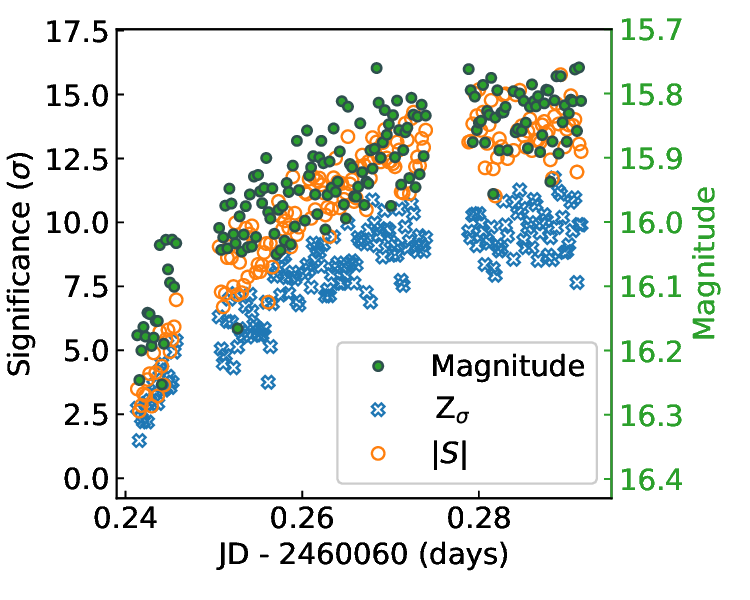}} 
    \caption{A comparison of $Z_\sigma$ (blue empty X's) and $\vert S\vert$ (orange empty circles) for a series of images 
    capturing a variable star in the FoV. The measured LAST magnitude of the star (green solid circles) is also 
    shown, with its value read on the right y-axis. For a stationary variable star, $\vert S\vert$ is dominant.}
\label{fig:var_star_S2_vs_Z2}
\end{figure}

For the variable source, we analyzed images capturing LINEAR~3506385, an eclipsing binary with a period of 
$\approx0.22\,$d and an average V-band magnitude of $15.96$ (\citealt{Drake+2014_CRTS_PeriodicVariables}). In the 
selected data, the observations cover a brightening phase of the binary. We thus use one of the first images as the 
reference image and perform subtraction of subsequent images,
registered to the reference image. The standard deviation in the astrometric solution of 
the binary's Right Ascension and Declination coordinates within the data is about $80\,$mas.
As the binary brightens over time, both statistics increase, whereas $|S|$ leads to overall higher significance than 
$Z_\sigma$ (Figure~\ref{fig:var_star_S2_vs_Z2}). This result is consistent with the interpretation of a variable  
stationary source. 

\begin{figure*}
    \centerline{\includegraphics[trim={0 0 0 0},clip,width=0.8\textwidth]{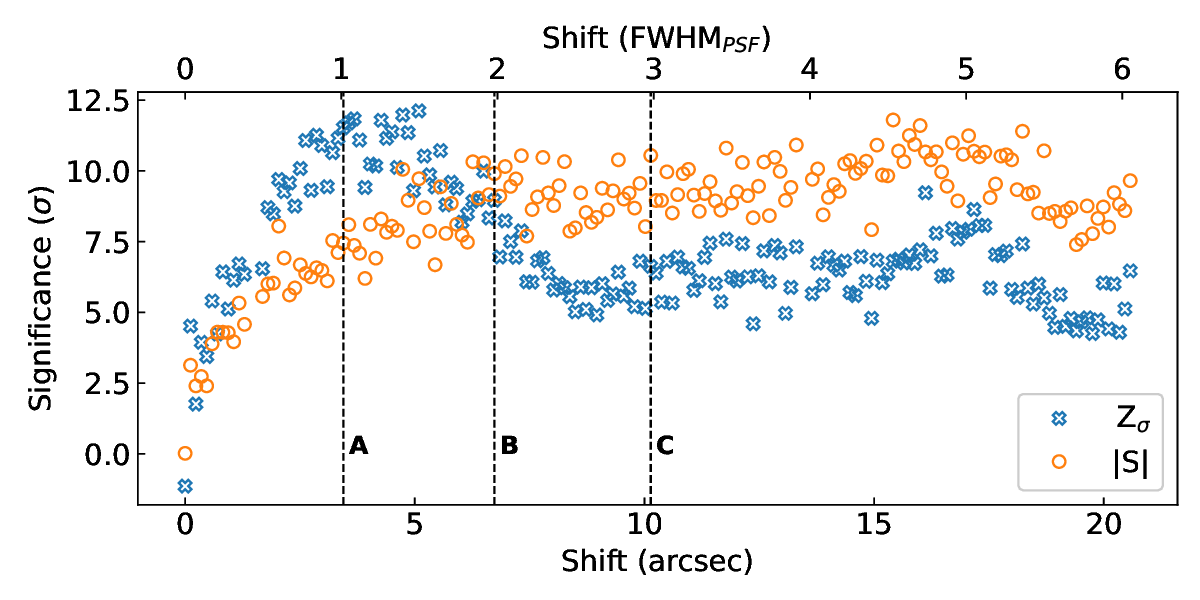}}
    \centerline{\includegraphics[trim={0 0 0.0 0},clip,width=0.75\textwidth]{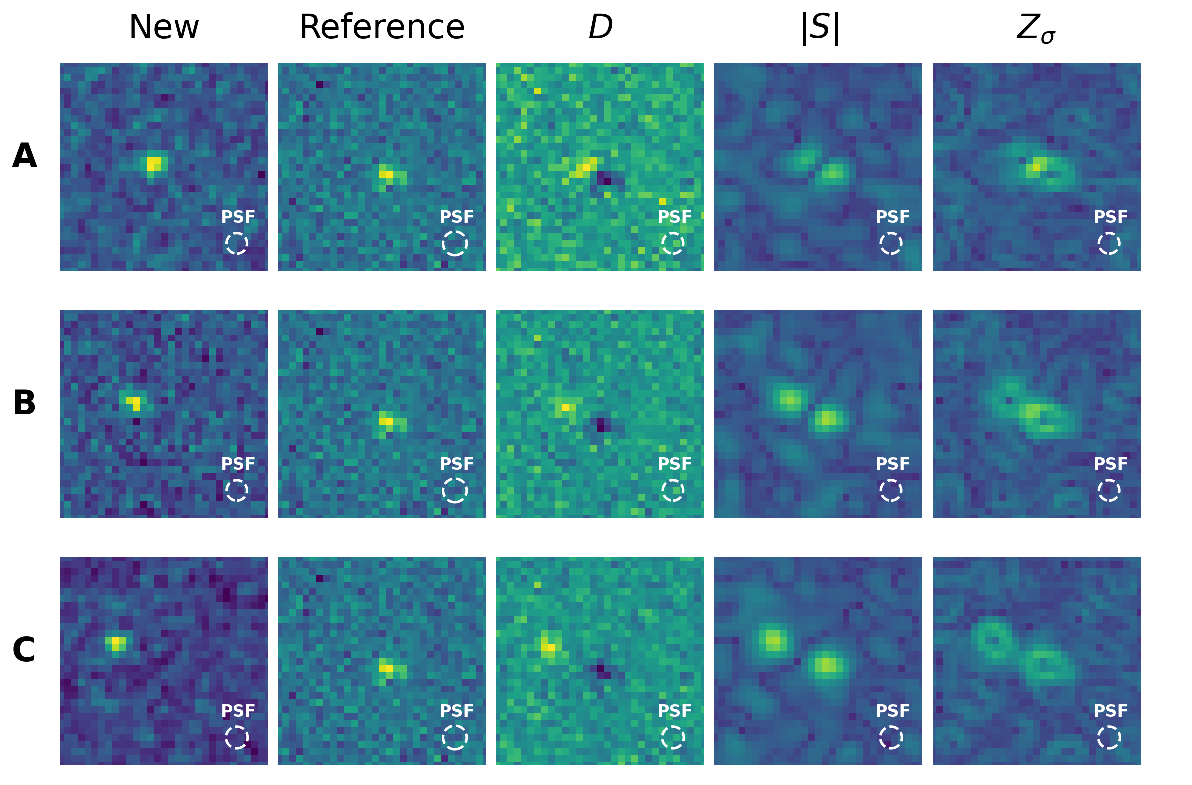}} 
    \caption{{\bf Upper panel:} A comparison of $Z_\sigma$ (blue empty X's) and $|S|$ (orange empty circles) for a 
    series of images capturing a moving source (asteroid~3832 Shapiro). The images are taken sequentially within one 
    hour during the same night. The first image serves as the reference image. Each sequential new image thus 
    represents a source moving away from its initial position. The distance, that the asteroid has crossed since the 
    time of the reference image, is shown as a shift in arcseconds and in the FWHM of $P_r$. Three selected 
    measurements, where the shift is closest to a multiple of FWHM$_{\text{PSF}}$, are marked by gray dashed vertical 
    lines and designated by the letters A, B, and C. {\bf Lower panel:} For each of the three marked measurements 
    (A, B, C), a set of cutouts of the new, reference, $D$, $|S|$, and $Z_\sigma$ images around the initial asteroid 
    position are shown. For the $|S|$ and $Z_\sigma$ cutouts, the minimum and maximum values of the colormap are 
    fixed to $-5\sigma$ and $13\sigma$, respectively. For each cutout, the dashed circle shows the FWHM of the PSF of 
    the respective image. For $|S|$ and $Z_\sigma$, the shown PSF is that of the $D$ image. Within a shift of 
    $\lesssim2\,$FWHM$_{\text{PSF}}$, $Z_\sigma > |S|$. Once the asteroid positions in the reference and new images 
    are clearly separated, $|S|$ overtakes $Z_\sigma$.}
\label{fig:asteroid_S2_vs_Z2}
\end{figure*}

As a moving source example, we analyzed images capturing the asteroid~3832 (Shapiro). The asteroid's motion 
between two successive images is about $\approx0.3$\,arcsec, whereas the LAST pixel scale is $1.25$\,arcsec
\,pix$^{-1}$.
We used the first image as a reference image so that the time series of subsequent registered images represents a source moving further away from its initial position.
Figure~\ref{fig:asteroid_S2_vs_Z2} shows $|S|$ and $Z_\sigma$ 
as functions of the asteroid's motion relative to the reference image. The results clearly show that for small 
translations, when the shift is smaller than the PSF size, $Z_\sigma$ yields a larger significance than $|S|$.
Once the asteroid has moved far enough to clearly separate it in the new image from itself in the 
reference image, it effectively becomes a transient, and $|S|$ overtakes $Z_\sigma$.

\section{Code}
\label{sec:Code}

The translient algorithm was implemented both as Python and MATLAB codes.
The Python code includes a core functionality that receives
the new image, a reference image, and their PSFs. The new and reference images should be flux-matched and registered.
The core function returns $\vec{\widehat{z}}$ data product.
This function, along with the likelihood functions and code to generate the plots in this paper, are available from GitHub\footnote{\url{https://github.com/ofersp/translient}}.

The MATLAB code is available as part of the AstroPack/MAATv2 package available from 
GitHub\footnote{\url{https://github.com/EranOfek/AstroPack}}.
The MATLAB code includes a core function\footnote{Function name {\tt imUtil.properSub.translient}},
a ZOGY plus \textsc{Translient} class\footnote{Implemented in the {\tt AstroZOGY} class.},
and many high-level functions that allow the construction of the PSF, registering the images, 
measure the zero points, apply proper subtraction, and more.
The image subtraction utilities are also described, with examples, in the AstroPack live wiki 
page\footnote{\url{https://github.com/EranOfek/AstroPack/wiki/imProc.sub} and \url{https://github.com/EranOfek/AstroPack/wiki/AstroZOGY}}.

\section{Discussion}
\label{sec:discussion}
We have presented a novel extension to the work of \cite{Zackay+2016_ZOGY_ImageSubtraction} to detect and measure 
transients that result from a point source undergoing translation given a pair of temporally separated images of the 
event. For the case of a pure translation (no flux change), and when the 
translation is smaller than the PSF size, we derive an optimal statistic that can detect such translients for any, a 
priori unknown, direction of motion. We show, using simulated examples, that our derived statistic, when applied to 
moving sources, reaches a higher true-positive detection rate at all false-positive detection rates, compared to the 
\textit{proper subtraction} statistic of \cite{Zackay+2016_ZOGY_ImageSubtraction}. This is the case at various levels 
of background noise and for various small, compared to the PSF, non-zero translations.

When dealing with transient detection,
this method should be used together with the proper image subtraction (ZOGY) method.
While ZOGY is optimal for flux variation, \textsc{translient} is optimal for motion.
Therefore, the combination of the two methods allows one to discern between variability (or transient detection) 
and motion (e.g., due to astrometric scintillation noise).
Specifically, calculating both statistics and comparing their significance allows us to tell
which model better explains the flux residuals in the subtraction image - is it variability or motion?
We have demonstrated this capability using real telescope images of an asteroid and a variable source.
We argue that this is a powerful statistic to distinguish between transients/variables and 
source motion due to registration errors, scintillation noise, color-refraction, or astrophysical motion. 

For the case where the transient detection results from both a change in position and a 
change in flux, we additionally derive the likelihood function for observing a particular difference image (shown to 
sufficiently capture all the necessary information) which is then used to disambiguate changes in flux and position. 
Using simulated examples, we have studied the behavior of this likelihood function, when the translation is either 
larger or smaller than the PSF diameter, and using three different parameterizations of the likelihood. We show that
for large translations, this likelihood function can distinguish between motion and flux variation in all cases.
For translations smaller than the PSF, a degeneracy (or strong dependence) of the parameters occurs in the likelihood 
in some of the parameterizations. Using a linearized version of the likelihood, we have found a reduced set of non-
degenerate parameters which are preferable for discrimination between translation and flux variation in the case of 
small translations.

Our method is unique in that it enables the detection of translating point sources in crowded fields and on top of a 
complex background for which no good models exist. It is derived from a well-defined generative process, is 
essentially non-parametric, and is optimal in the case of small pure translations. 
Particular cases where the newly developed detector method may find use are (but not limited to):
(i) reducing the amount of false alarms, due to motion, in a transient detection algorithm implementation;
and (ii) measuring the motion of astrophysical sources on top of a complicated background.

\acknowledgments

E.O.O. is grateful for the support of
grants from the 
Willner Family Leadership Institute,
André Deloro Institute,
Paul and Tina Gardner,
The Norman E Alexander Family M Foundation ULTRASAT Data Center Fund,
Israel Science Foundation,
Israeli Ministry of Science,
Minerva,
BSF, BSF-transformative, NSF-BSF,
Israel Council for Higher Education (VATAT),
Sagol Weizmann-MIT,
Yeda-Sela, and the
Rosa and Emilio Segre Research Award.
This research is supported by the Israeli Council for Higher Education (CHE) via the Weizmann Data Science Research Center, and by a research grant from the Estate of Harry Schutzman.

\clearpage
\appendix

\section{Sufficiency of the difference image statistic $\widehat{D}_T$}
\label{sec:D_suff}

Here we show that the difference image $\widehat{D}_T$ defined in \S\ref{sec:discrim} is a \textit{sufficient 
statistic} for determining the model parameters $\alpha_r$, $\alpha_n$, $\vec{q}$ and $\vec{p}$. This is due to the 
fact that, given the model parameters, the PSF kernels $P_r$ and $P_n$ and the observational noise variances 
$\sigma_r^2$ and $\sigma_n^2$ (which we assume to be known), there is a one to one mapping between the difference 
image $\widehat{D}_T$ (defined by Equation~\ref{eq:D})
\begin{equation}
    \widehat{D}_T \equiv \widehat{P}_r\widehat{N} - \widehat{P}_n \widehat{R} + \widehat{P}_n\widehat{P}_r(\alpha_r\widehat{\delta}_{\vec{q}} - \alpha_n\widehat{\delta}_{\vec{p}}),
\end{equation}
and the \textit{proper subtraction} image $\widehat{D}$ defined by Equation~(A21) of  \cite{Zackay+2016_ZOGY_ImageSubtraction}:
\begin{equation}
    \widehat{D} \equiv \frac{\sqrt{\sigma_r^2+\sigma_n^2}\left(\widehat{P}_r\widehat{N} - \widehat{P}_n \widehat{R}\right)}{\sqrt{\sigma_n^2|\widehat{P}_r|^2 + \sigma_r^2|\widehat{P}_n|^2}},
\end{equation}
and therefore, $P(\widehat{D}_T | \alpha_r, \alpha_n, \vec{q}, \vec{p}) = P(\widehat{D} | \alpha_r, \alpha_n, \vec{q}, \vec{p})$. Moreover, \cite{Zackay+2016_ZOGY_ImageSubtraction} show that the proper subtraction image $\widehat{D}$ is a sufficient statistic for determining the parameters of a family of generative processes which include the generative process of Equations \ref{eq:model_R} and \ref{eq:model_N}. Hence, applying the result of \cite{Zackay+2016_ZOGY_ImageSubtraction} to our model parameters, we obtain
\begin{eqnarray}
P(N, R | \alpha_r, \alpha_n, \vec{q}, \vec{p}) &=& P(\widehat{D} | \alpha_r, \alpha_n, \vec{q}, \vec{p})h(R, N) \\
&=& P(\widehat{D}_T | \alpha_r, \alpha_n, \vec{q}, \vec{p})h(R, N),
\end{eqnarray}
for some non-negative function $h$ which does not depend on the model parameters. This, according to the Fisher-Neyman factorization theorem \citep{Fisher+1922RSPTA_SufficientStatistics}, characterizes $\widehat{D}_T$ as a \textit{sufficient statistic} with respect to the parameters $\alpha_r$, $\alpha_n$, $\vec{q}$, and $\vec{p}$. 
In practice, this means that 
\begin{eqnarray}
\log P(\widehat{D}_T | \alpha_r, \alpha_n, \vec{q}, \vec{p}) = \log P(N, R | \alpha_r, \alpha_n, \vec{q}, \vec{p}) + \mathrm{const},
\end{eqnarray}
where the additive constant does not depend on the model parameters.

\bibliography{papers.bib}
\bibliographystyle{apalike}

\end{document}